\documentclass[aps,prd,twocolumn,superscriptaddress,nofootinbib]{revtex4-1}

\usepackage{amsfonts}
\usepackage{psfrag}
\usepackage{graphicx}
\usepackage{grffile}
\usepackage{cancel}
\usepackage{amsmath}
\usepackage{natbib}
\usepackage{xr}
\usepackage{hyperref}
\usepackage{verbatim}
\usepackage{epstopdf}
\usepackage{subcaption}
\usepackage{multirow}
\usepackage{commath}
\usepackage{xcolor}
\hypersetup{    
  colorlinks      = true,
  linkcolor       = {blue},
  linkbordercolor = {white},
  citecolor       = {blue},
  citebordercolor = {white},
  urlcolor        = {blue},
  urlbordercolor  = {white},
}

\usepackage{soul}

\usepackage{booktabs,dcolumn,caption}

\newcolumntype{d}[1]{D{.}{.}{#1}}

\RequirePackage[normalem]{ulem} 
\RequirePackage{color}\definecolor{RED}{rgb}{1,0,0}\definecolor{BLUE}{rgb}{0,0,1} 
\providecommand{\DIFaddbegin}{} 
\providecommand{\DIFaddend}{} 
\providecommand{\DIFdelbegin}{} 
\providecommand{\DIFdelend}{} 
\providecommand{\DIFaddbeginFL}{} 
\providecommand{\DIFaddendFL}{} 
\providecommand{\DIFdelbeginFL}{} 
\providecommand{\DIFdelendFL}{} 
\newcommand{\DIFscaledelfig}{0.5}
\RequirePackage{settobox} 
\RequirePackage{letltxmacro} 
\newsavebox{\DIFdelgraphicsbox} 
\newlength{\DIFdelgraphicswidth} 
\newlength{\DIFdelgraphicsheight} 
\LetLtxMacro{\DIFOincludegraphics}{\includegraphics} 
\newcommand{\DIFaddincludegraphics}[2][]{{\color{blue}\fbox{\DIFOincludegraphics[#1]{#2}}}} 
\newcommand{\DIFdelincludegraphics}[2][]{
\sbox{\DIFdelgraphicsbox}{\DIFOincludegraphics[#1]{#2}}
\settoboxwidth{\DIFdelgraphicswidth}{\DIFdelgraphicsbox} 
\settoboxtotalheight{\DIFdelgraphicsheight}{\DIFdelgraphicsbox} 
\scalebox{\DIFscaledelfig}{
\parbox[b]{\DIFdelgraphicswidth}{\usebox{\DIFdelgraphicsbox}\\[-\baselineskip] \rule{\DIFdelgraphicswidth}{0em}}\llap{\resizebox{\DIFdelgraphicswidth}{\DIFdelgraphicsheight}{
\setlength{\unitlength}{\DIFdelgraphicswidth}
\begin{picture}(1,1)
\thicklines\linethickness{2pt} 
{\color[rgb]{1,0,0}\put(0,0){\framebox(1,1){}}}
{\color[rgb]{1,0,0}\put(0,0){\line( 1,1){1}}}
{\color[rgb]{1,0,0}\put(0,1){\line(1,-1){1}}}
\end{picture}
}\hspace*{3pt}}} 
} 
\LetLtxMacro{\DIFOaddbegin}{\DIFaddbegin} 
\LetLtxMacro{\DIFOaddend}{\DIFaddend} 
\LetLtxMacro{\DIFOdelbegin}{\DIFdelbegin} 
\LetLtxMacro{\DIFOdelend}{\DIFdelend} 
\DeclareRobustCommand{\DIFaddbegin}{\DIFOaddbegin \let\includegraphics\DIFaddincludegraphics} 
\DeclareRobustCommand{\DIFaddend}{\DIFOaddend \let\includegraphics\DIFOincludegraphics} 
\DeclareRobustCommand{\DIFdelbegin}{\DIFOdelbegin \let\includegraphics\DIFdelincludegraphics} 
\DeclareRobustCommand{\DIFdelend}{\DIFOaddend \let\includegraphics\DIFOincludegraphics} 
\LetLtxMacro{\DIFOaddbeginFL}{\DIFaddbeginFL} 
\LetLtxMacro{\DIFOaddendFL}{\DIFaddendFL} 
\LetLtxMacro{\DIFOdelbeginFL}{\DIFdelbeginFL} 
\LetLtxMacro{\DIFOdelendFL}{\DIFdelendFL} 
\DeclareRobustCommand{\DIFaddbeginFL}{\DIFOaddbeginFL \let\includegraphics\DIFaddincludegraphics} 
\DeclareRobustCommand{\DIFaddendFL}{\DIFOaddendFL \let\includegraphics\DIFOincludegraphics} 
\DeclareRobustCommand{\DIFdelbeginFL}{\DIFOdelbeginFL \let\includegraphics\DIFdelincludegraphics} 
\DeclareRobustCommand{\DIFdelendFL}{\DIFOaddendFL \let\includegraphics\DIFOincludegraphics} 

\begin{document}

\title{Tidal disruption of  solitons in self-interacting ultralight axion dark matter}

\date{Received 23 May 2022; accepted 6 June 2022}

\author{Noah Glennon}
\email{nglennon@wildcats.unh.edu}
\affiliation{Department of Physics and Astronomy, University of New Hampshire, Durham, New Hampshire 03824, USA}

\author{Ethan~O.~Nadler}
\email{enadler@carnegiescience.edu}
\affiliation{Carnegie Observatories, 813 Santa Barbara Street, Pasadena, CA 91101, USA}
\affiliation{Department of Physics $\&$ Astronomy, University of Southern California, Los Angeles, CA, 90007, USA}

\author{Nathan Musoke}
\email{nathan.musoke@unh.edu}
\affiliation{Department of Physics and Astronomy, University of New Hampshire, Durham, New Hampshire 03824, USA}

\author{Arka Banerjee}
\email{arka@iiserpune.ac.in}
\affiliation{Department of Physics, Indian Institute of Science Education and Research,
Homi Bhabha Road, Pashan, Pune 411008, India}
\affiliation{Fermi National Accelerator Laboratory, Cosmic Physics Center, Batavia, IL 60510, USA}

\author{Chanda Prescod-Weinstein}
\email{chanda.prescod-weinstein@unh.edu}
\affiliation{Department of Physics and Astronomy, University of New Hampshire, Durham, New Hampshire 03824, USA}

\author{Risa~H.~Wechsler}
\email{rwechsler@stanford.edu}
\affiliation{Kavli Institute for Particle Astrophysics and Cosmology and Department of Physics, Stanford University, Stanford, CA 94305, USA}
\affiliation{SLAC National Accelerator Laboratory, Menlo Park, CA 94025, USA}

\begin{abstract}
Ultralight axions (ULAs) are promising dark matter candidates that can have a distinct impact on the formation and evolution of structure on nonlinear scales relative to the cold, collisionless dark matter (CDM) paradigm. However, most studies of structure formation in ULA models do not include the effects of self-interactions, which are expected to arise generically. Here, we study how the tidal evolution of solitons is affected by ULA self-interaction strength and sign. Specifically, using the pseudospectral solver \texttt{UltraDark.jl}, we simulate the tidal disruption of self-interacting solitonic cores as they orbit a $10^{11}~M_{\mathrm{\odot}}$ Navarro-Frenk-White CDM host halo potential for a range of orbital parameters, assuming a fiducial ULA particle mass of $10^{-22}\mathrm{eV}$. We find that repulsive (attractive) self-interactions significantly accelerate (decelerate) soliton tidal disruption.  We also identify a degeneracy between the self-interaction strength and soliton mass that determines the efficiency of tidal disruption, such that disruption timescales are affected at the $\sim 50\%$ level for variations in the dimensionless ULA self-coupling from $\lambda=-10^{-92}$ to $\lambda=10^{-92}$.
\end{abstract}

\maketitle

\section{Introduction}

Evidence from the cosmic microwave background (CMB), galactic rotation curves, galaxy clustering, the Lyman-$\alpha$ forest, and gravitational lensing indicates that the majority of matter in the universe is dark \cite{Ade2015,Bertone2005,Bertone2018,Freese2009,Chabanier:2019eai}. Dark matter is usually assumed to be cold and collisionless (CDM; \cite{Buckley2018, Armendariz2014}). However, an enormous range of particle dark matter models that are compatible with current cosmological, collider, and direct detection experiments break these assumptions in detail, yielding unique astrophysical signatures that will be probed by upcoming observational facilities~\cite{Drlica-Wangner2019,Bechtol2019,Gezari:2022rml,Valluri:2022nrh}.

An increasingly popular class of dark matter models feature scalar fields with a shift symmetry~\cite{Braine2019, Ouellet2019, Chaudhuri2018}. This class of models includes candidates like the QCD axion, which was originally theorized to solve the Strong CP Problem~\cite{Peccei1977}, and ultralight axions (ULAs), which have masses on the order of $\sim 10^{-22}\mathrm{eV}$ or smaller. Axion-like particle (ALP) models that are motivated by string theory suggest there may be many different axion particles with a wide range of masses.  For example, the axiverse hypothesis suggests that there may be many ALPs with different masses in decades from $\sim 1$ eV down to $10^{-33}~\text{eV}$~\cite{Arvanitaki2010,Marsh2016}.  In ULA models, the particles are bosonic and under certain circumstances can behave coherently as a Bose--Einstein condensate (BEC) due to their extremely high occupation numbers~\cite{Seidel1990,Kolb1993,Kolb1994,Sikivie2009,Guth2015}. Meanwhile, the de Broglie wavelength for ULAs is roughly on the kiloparsec scale \cite{Hui2017}, and the clustering of these particles is suppressed on smaller scales. This leads to a cutoff in the halo mass function and yields dark matter cores rather than cusps on scales corresponding to dwarf galaxies~\cite{Schive2014,Schive2014_2,Marsh2015a,Mocz2017,Veltmaat2019,Schwabe2016,Schwabe:2020eac,Schwabe2021,Hu2000}. Thus, the existence of ULA fields with conspicuous astrophysical signatures is theoretically motivated.

Below the QCD scale, the potential for axion or axion-like particles often takes the form
\begin{equation}
    V(\phi)=\Lambda^4\left(1-\cos\left(\phi/f_a\right)\right) \text{.}\label{eq:Vphi}
\end{equation}
Here, $\phi$ is the axion field, $\Lambda
\approx~0.1$ GeV, and $f_a$ is the Peccei--Quinn symmetry breaking scale or axion decay constant~\cite{Marsh2016, Dine1983}.  By expanding the potential, we see that axions generically undergo self-interactions, with leading order term $\phi^4$.  It is important to account for this term because, as shown in~\cite{Chavanis2016, Glennon2020}, even small self-interactions can have large effects on the the dynamics of ULA dark matter.  Many axion-like dark matter models ignore potential self-interactions because they are constrained to be very small~\cite{Fan2016, Chavanis2016}. 
For example, ULA models in the free-field limit are sometimes referred to as fuzzy dark matter (FDM) models~\citep[][]{Hu2000,Schive2014}. Although there are a variety of astrophysical constraints on the FDM mass (e.g., \cite{Irsic:2017yje,Armengaud:2017nkf,Corasaniti:2016epp,Safarzadeh:2019sre,Schutz:2020jox,Rogers:2020ltq,DES:2020fxi,Dalal:2022rmp}), none of these analyses incorporate the effects of potential ULA self-interactions.\footnote{However, see \cite{Shapiro:2021hjp,Dawoodbhoy:2021beb} for analytic studies and idealized simulations of ULAs with repulsive self-interactions in the Thomas-Fermi regime, and see \cite{Chakrabarti:2022owq} for joint constraints on the ULA mass and self-coupling from the enclosed mass profile of M87.} However, as seen in prior work, even a small self-interaction can greatly impact the resulting dynamics because axion-like particle densities can be very large~\cite{Chavanis2011,Chavanis2016,Desjacques2017}.  For example, any attractive self-interaction ensures that there is a maximum mass that a bound dark matter state (or ``soliton'') can have before collapsing into a black hole, and also allows for oscillating or exploding solitons~\cite{Chavanis2016, Glennon2020}.  There are are also compelling reasons to study ULAs with repulsive self-interactions~\cite{Fan2016}.

The tidal evolution of gravitationally bound dark matter subhalos as they orbit within a larger host halo dictates the properties of small-scale structure at late times. In turn, this physics has important consequences for the interpretation of observational probes of low-mass subhalos. In particular, studies leveraging strong gravitational lensing \cite{Gilman:2019vca}, the Milky Way satellite galaxy population \cite{DES:2019ltu}, and stellar stream perturbations \cite{Banik:2019cza} have recently gained sensitivity to subhalos as small as $\sim 10^8\ M_{\mathrm{\odot}}$. At fixed particle mass, characteristic ULA effects including soliton sizes \cite{Schive2014_2} and gravitational heating due to wave interference \cite{Hui2017,Dalal2022} become \emph{larger} with decreasing subhalo mass and velocity dispersion. Precise theoretical predictions for the evolution of these systems in realistic ULA models, including self-interactions, are therefore timely.

Several previous studies have considered the tidal evolution of ULA subhalos and solitons. Specifically, \cite{Du2018} simulates a soliton orbiting a central potential using a pseudospectral solver to evolve the Schr\"{o}dinger-Poisson equation. In the absence of self-interactions, these authors find that, once the soliton drops below a critical fraction of the host's average density within the orbital radius, tidal stripping results in runaway disruption. Meanwhile, \cite{Schive:2019rrw} show that the presence of an outer CDM-like halo profile surrounding the soliton can make the soliton significantly more resilient to tidal disruption, and \cite{Li:2020ryg} propose that an eigenmode analysis provides insight into this process. None of these studies consider the coupled effects of ULA self-interactions and tidal stripping, motivating our study. Furthermore, cosmological simulations of ULAs (e.g., \cite{Schive2014,Zhang:2018ghp,May:2021wwp,Schwabe:2020eac,Schwabe2021}) currently lack the resolution to resolve solitons' detailed tidal evolution within larger host halos, and no cosmological simulations to date include the effects of ULA self-interactions.

Here, we present simulations of ULA dark matter using a version of \texttt{UltraDark.jl}~\cite{ultradark}, which is capable of accounting for self-interactions.  We attempt to characterize the effects of self-interactions on ULA dark matter as it tidally disrupts.  We find that there is a degeneracy between the self-interaction strength and soliton mass in determining the disruption time.  This work extends earlier results looking at tidal disruption in ULA dark matter~\cite{Du2018}, introducing the possibility of self-interactions and showing that they can have a significant impact on the disruption times of solitons.

This paper is organized as follows. In Section \ref{sec:ULA}, we present our ULA model and the physical setup we implement in our soliton simulations. In Section~\ref{sec:analytic}, we estimate the tidal radii of the solitons considered in our simulations to qualitatively assess how efficiently these systems will tidally disrupt. In Section~\ref{sec:sim_description}, we discuss the pseudospectral solver used for our simulations, \texttt{UltraDark.jl}, and the initial conditions for our simulations. In Section~\ref{sec:results}, we discuss the results of our simulations, including how soliton disruption time depends on self-interaction strength and sign, soliton mass, and orbital parameters.  We show that self-interaction strength and sign and soliton mass influence the disruption time in a degenerate fashion.  In Section~\ref{sec:discussion}, we discuss the importance of our work and its main caveats.  Lastly, in Section~\ref{sec:conclusions}, we summarize our results.  Throughout, we work in units with $c= G = 1$.

\section{ultralight Axion Model and Physical Setup}
\label{sec:ULA}

Our ULA model consists of a classical field minimally coupled to gravity.  Considering only the leading order self-interaction term from Eq.~\ref{eq:Vphi}, the action takes the form
\begin{equation}
    S = \int d^4 x \sqrt{-g} \left[\frac{1}{2} g^{\mu\nu} \partial_{\mu}\phi\partial_{\nu}\phi-\frac{1}{2}m^2\phi^2-\frac{\lambda}{4}\phi^4\right]. 
\end{equation}
Here, $\phi$ is the scalar field, $m$ is the mass of the field, and $\lambda$ is the dimensionless self-coupling.  We obtain the equations of motion by writing the real scalar field $\phi$ in terms of a complex field $\psi$,
\begin{equation}
    \phi = \frac{\hbar}{\sqrt{2m}}\left(\psi e^{-i m t/\hbar} + \psi^* e^{i m t/\hbar}\right).
\end{equation}
The equations of motion in the Newtonian gauge are the Gross-Pitaevskii-Poisson (GPP) equations, 
\begin{equation}
	i \hbar \dot{\psi} = -\frac{\hbar^2}{2m} \nabla^2 \psi + m \Phi \psi + \frac{\hbar^3 \lambda}{2m^3} \abs{\psi}^2 \psi
	\label{eq:GPeqn}
\end{equation}
and
\begin{equation}
	\nabla^2 \Phi = 4\pi G m \abs{\psi}^2.
	\label{eq:Peqn}
\end{equation}
Here, $\Phi$ is the gravitational potential and $\psi$ is the ULA field.  A more detailed derivation of the equations of motion can be found in~\cite{Kirkpatrick2020}.  In the following analysis, using a particle mass of $m=10^{-22}~\mathrm{eV}$, the self-interaction strength is parameterized by  a dimensionless variable $\tilde{\kappa}$, defined as 
\begin{equation}
    \tilde{\kappa} \equiv 2.1 \times 10^{92} \lambda
\end{equation}
Table~\ref{tab:param_ranges} details the relevant units and ranges of ULA parameter values in our study.

There are different approaches to constraining the self-interaction strength of ULAs~\cite{Fan2016, Chavanis2016, Li:2013nal}.  Different methods can lead to constraints that have large order of magnitude differences. Here, we use a dimensionless coupling on the order of $\mathcal{O}(10^{-92})$ that falls within the bounds found in~\cite{Li:2013nal}. Although the self-interaction strength is predicted to be very small, the effects of the self-interactions are governed by the self-interaction strength times the phase-space density of axions in the region of interest~\cite{Desjacques2017}. In particular, the non-linear self-interaction term in the adimensional GPP equations (see Eq.~\ref{eq:GPeqn}) is given by $\lambda \abs{\psi}^2 \psi$; since the density is given by $\rho = m \abs{\psi}^2$, the non-linear term is proportional to $\lambda \rho \psi$.  Since the denisty of ULAs can be very large in the inner regions of solitons, the non-linear effects of the self-interactions may be important despite the self-interaction strengths being small.  In this work, we consider attractive and repulsive self-interactions with similar magnitudes because there are numerical difficulties when self-interactions are large.

The term soliton refers in this work to the localized dark matter that is a Bose--Einstein condensate.  In ULA models, dark matter halos are composed of a solitonic core with an NFW outer region~\cite{Schive2014,Marsh2015a}.  In our simulations, we consider a soliton orbiting a more massive host to mimic the central region of a dwarf galaxy falling into a larger system.  For most of this work, we  simulate the solitonic core being tidally disrupted without the NFW region in order to isolate the effects of self-interactions on the core disruption time. We assume a fiducial ULA mass of $10^{-22}~\mathrm{eV}$.

\section{Analytic Estimates of Soliton Tidal Radii}
\label{sec:analytic}

In this Section, we use the physical setup described in Section \ref{sec:ULA} to estimate the tidal radii of solitons orbiting a host halo. The tidal radius represents the distance within which the soliton's gravity dominates over the host's tides, and thus provides an estimate of which regions are protected from tidal disruption. Note that, due to the wave-like nature of ULAs, all regions of the soliton can potentially be tidally stripped regardless of the tidal radius \cite{Hui2017,Li:2020ryg}. However, the tidal radius still provides a useful means to qualitatively assess how efficiently disruption can proceed for a given solitonic and orbital configuration.

We estimate the tidal radii of solitons in circular orbits using several related definitions presented in \cite{vandenBosch2017}. First, consider the tidal radius $r_{t,1}$~\cite{vandenBosch2017}, 
\begin{equation}
    r_{t,1} = R \left[\frac{m(r_{t,1})/M(R)}{2-\frac{d\ln M}{d\ln R}|_R}\right]^{1/3},
\end{equation}
where $R$ is the separation between the centers of the soliton and the host, $m(r_{t,1})$ is the mass of the soliton, and $M(R)$ is the mass of the host within the orbital radius. This result only assumes that the soliton and host have extended mass profiles, and ignores effects like centrifugal force. The host mass within the orbital radius can be found using
\begin{equation}
    M(R) = 4\pi\rho_0 R^3_s\left[\ln\left(\frac{R_s+R}{R_s}\right)+\frac{R_s}{R_s+R}-1\right]
    \label{eq:rt3}
\end{equation}
where $\rho_0$ is a density parameter and $R_s$ is the scale radius of the central NFW potential.    In our simulations, we adopt $c = 20$, $r_{vir} = 95~\mathrm{kpc}$, and $M(r_{vir}) = 10^{11} ~M_\odot$.  This gives a density parameter of $\rho_0 = 3.5 \times 10^{7}~M_{\odot}/\mathrm{kpc}^3$.  Across the range of values of $M$ and $R$ we consider, $d\ln M/d\ln R$ is roughly constant, with a typical value of about 0.46.  According to Eq.~\ref{eq:rt3}, our solitons' tidal radii are then on the order of 8 kpc.

We can alternatively define the tidal radius as \citep{vandenBosch2017}
\begin{equation}
    r_{t,2} = R \left[\frac{m(r_{t,2})/M(R)}{3-\frac{d\ln M}{d\ln R}|_R}\right]^{1/3},
\end{equation}
which accounts for centrifugal force, or using  
\begin{equation}
    r_{t,3} = R\left[\frac{m(r_t)}{M(R)}\right]^{1/3},
\end{equation}
which corresponds to the radius where the frequency of tidal forces applied by the host matches the internal motion of the soliton. These definitions satisfy $r_{t,2} < r_{t,1} < r_{t,3}$ for the configurations we simulate, implying that $r_{t,1}\approx 8 \rm{kpc}$ remains a reasonable estimate.

The solitons we initialize have core radii of roughly $2~\text{kpc}$.  This is smaller than the tidal radius by a factor of a few, indicating that tidal disruption will not occur immediately, but that it can plausibly occur once the core radius expands sufficiently due to tidal stripping.  Importantly, self-interactions influence the dynamical evolution of our solitons in a manner that is not easily captured by tidal radius calculations. Qualitatively, we expect attractive (repulsive) self-interactions to make solitons less susceptible (more susceptible) to disruption and therefore to effectively increase (decrease) the tidal radius, but the extent to which the tidal radius alone can accurately characterize the disruption of self-interacting solitons is unclear. Our simulations are therefore crucial to quantify how self-interactions affect tidal disruption.

We note that, in a previous study of soliton tidal disruption without self-interactions,~\cite{Du2018} found that the tidal radius of a soliton depends only on the ratio of its core density to the average density of the host within the soliton's orbit.  These authors found that if this ratio is less than 4.5, the tidal radius of the soliton is smaller than its core radius, leading to rapid disruption: as the core loses mass, its density drops and it expands further in a runaway process. In most of our simulations, the initial density ratio is $\approx 50$, which ensures that $>95\%$ of the soliton's mass is contained within the tidal radius at first~\cite{Du2018}, protecting the soliton from immediate runaway disruption due to the host's tides. The sharp transition to runaway disruption at low density ratios motivates our choice to simulate a narrow range of soliton masses with initial density ratios above the critical threshold.

\section{Simulation Description}
\label{sec:sim_description}

\subsection{Numerical Methods}

We perform our simulations using \texttt{UltraDark.jl}, a code that can be used to simulate ultralight dark matter~\cite{ultradark}, with modifications to allow for self-interactions.  
\texttt{UltraDark.jl} uses a pseudospectral method to solve the GPP equations (Eq.~\ref{eq:GPeqn}--\ref{eq:Peqn}), similar to \texttt{PyUltraLight} and \texttt{PySiUltraLight}~\cite{Edwards2018,Glennon2020}.
This means that the linear and nonlinear operators in Eq.~\ref{eq:GPeqn} are computed in Fourier space and configuration space respectively.
In particular, a single step in the evolution with self-interactions is given by
\begin{equation}
    \begin{split}
    	\psi(\vec{x},t+h) 
    	=&
    	\exp\left[-\frac{ih}{2} \Phi(\vec{x},t+h)\right]
    	\\
    	& \times \exp\left[-\frac{ih\kappa}{2}\abs{\psi(\vec{x},t+h)}^2\right]
    	\\
    	& \times \mathcal{F}^{-1}\exp\left[-\frac{ih}{2}k^2\right]
    	\mathcal{F} \exp\left[-\frac{ih}{2} \Phi(\vec{x},t)\right]
    	\\
    	& \times \exp\left[-\frac{ih\kappa}{2}\abs{\psi(\vec{x},t)}^2\right] \psi(\vec{x},t),
	\end{split}
\end{equation}
where $h$ is a small time step, $\mathcal{F}$ is the Fourier transform, $\mathcal{F}^{-1}$ is the inverse Fourier transform, $k$ is the wavenumber in Fourier space, $\psi(\vec{x},t_i)$ is the field at the half step.
The gravitational potential is updated in phase space,
\begin{equation}
	\Phi(\vec{x},t+h) = \mathcal{F}^{-1}\left(-\frac{1}{k^2}\right)  \mathcal{F}4\pi \abs{\psi(\vec{x},t_i)}^2
	+ \Phi_{\text{ext}}\;,
\end{equation}
and a fixed background gravitational field $\Phi_{\text{ext}}$ is added.
A detailed description of how the ULA field with self-interactions and an external gravitational potential evolves over time can be found in~\cite{Glennon2020}.

There are some computational limitations associated with this method.
The primary constraint we encountered came from the so-called ``maximum velocity criterion.''
Specifically, when recasting the GPP to the Madelung equations, it is apparent that the velocity of the field is equal to the gradient of its phase~\cite{Woo:2008nn}
\begin{equation}
    \mathbf{v} = \frac{\hbar}{m} \nabla \arg(\psi)
    \;.
\end{equation}
However, information about this phase gradient is lost when $\psi$ is represented on a grid with finite resolution; the phase difference between neighboring cells is only known modulo $\pi$ and so the velocity is represented modulo
\begin{equation}
    v_{\mathrm{max}} = \frac{\hbar}{m} \frac{\pi}{\Delta x} \;,
    \label{vmax}
\end{equation}
where $\Delta x$ is the grid point spacing, set by the size of the simulation box and the resolution of the simulation.  Attempts to simulate velocities greater than $v_{\mathrm{max}}$ result in unphysical numerical artifacts and so \texttt{UltraDark.jl} terminates if velocities exceed $v_{\mathrm{mx}}/4$.

Since we use an FFT-based method to solve the equations of motion, it is natural to adopt periodic boundary conditions for our simulations, meaning that the simulation box is topologically equivalent to a torus.  This implies that angular momentum is not necessarily conserved if matter crosses the boundary of the box. This issue is mitigated simply by minimizing the amount of matter that crosses the boundary. We therefore use a box size that is large enough to comfortably encompass our solitons' orbits and stripped material, and small enough to yield reasonable maximum velocity criteria for these systems. To balance these factors, we use a box length that is about 320\% larger than the host's virial radius. We found that this box size does not compromise the maximum velocity criteria for any of our simulations while preventing a significant amount of matter from crossing the boundary of the box.

\subsection{Physical Setup}

We consider solitons on circular orbits around a central halo, which we model using a static, external gravitational potential corresponding to an Navarro-Frenk-White (NFW; \cite{Navarro:1996gj}) halo,
\begin{equation}
    \Phi_{\text{ext}}(r)
    = - \frac{M_{\text{host}}}{r} \frac{\ln(1 + c_\text{host} \frac{r}{r_{\text{vir}}})}{\ln(1+c_\text{host})-\frac{c_\text{host}}{1+c_\text{host}}}
    \,,
\end{equation}
where $\Phi_{\text{ext}}$ is the gravitational potential, $c_{\text{host}}$ is the host concentration, $M_{\text{host}}$ is the host mass, and $r_{\text{vir}}$ is the viral radius of the host. 
We simulate a host with mass $M_{\mathrm{host}} = 10^{11}\mathrm{M_\odot}$ and concentration $c_{\mathrm{host}} = 20$, roughly corresponding to the properties of halos that host the Large Magellanic Cloud (LMC) \cite{Erkal181208192,Shipp210713004} and similar galaxies. The LMC is known to host faint satellite galaxies (e.g., \cite{Kallivayalil}), which would be affected by ULA physics given our fiducial particle mass of $10^{-22}~\rm{eV}$ model (e.g., \cite{Hui2017}), adding to the astrophysical relevance of our simulations.

\setlength{\tabcolsep}{1pt}
\begin{table}[t!]
\centering
 \begin{tabular}{||c | c c ||} 
 \hline
 Parameter & Range & Definition \\ [0.45ex] 
 \hline\hline
 $m_a$ & $10^{-22} \mathrm{eV}$ & Particle mass\\
 $\tilde{\kappa}$ & -5--5 & Self-coupling strength \\ 
 $x_c$ & 0.7--1.1 & Orbital energy\\
 $r_\text{cir}$ & 67--105 kpc & Orbital radius\\
 $\eta$ & 1.0 & Circularity\\
 $M_{\mathrm{host}}$ & $10^{11}~M_\odot$ & Host mass\\
 $M_{\mathrm{sol}}$ & $1.04$--$1.13\times10^8~M_\odot$ & Soliton mass \\
 $c_{\mathrm{host}}$ & 20 & Host concentration\\
 \hline
 \end{tabular}
 \caption{The definitions and values of the parameters that enter our simulations. Orbital energy ($x_c$) and circularity ($\eta$) are defined according to Eq.~\ref{eq:energy}--\ref{eq:circularity}.}
 \label{tab:param_ranges}
\end{table}

Our simulations use a fiducial box length of 306 kpc and resolution of 512 grid points per side. They are initialized with a soliton, specified by its mass, position, velocity, phase, and density profile. The position and velocity are derived from the orbital energy $x_c$ and orbital circularity $\eta$, following~\cite{Ogiya2019}. Specifically, these parameters are defined as \cite{Ogiya2019}
\begin{align}
    x_c \equiv \frac{r_c(E)}{r_{\mathrm{vir}}}, \label{eq:energy}\\
    \eta \equiv \frac{L}{L_c(E)}.\label{eq:circularity}
\end{align}
Thus, $x_c$ corresponds to radius of a circular orbit with total energy $E$, in units of the host's virial radius, and $\eta$ is the ratio of the orbital angular momentum $L$ to that for a circular orbit of total energy $E$. Thus, $\eta = 0$ ($\eta=1$) corresponds to a radial (circular) orbit; throughout, we simulate solitons on circular orbits to minimize the impact of the maximum velocity criterion. Given this assumption, $x_c$ effectively parameterizes the soliton's distance from the host center, which we often quote in units of the host halo's virial radius. We summarize the range of $x_c$ we use, which avoids violating the maximum velocity criterion (Eq.~\ref{vmax}), in Table~\ref{tab:param_ranges}.

\begin{figure}[b!]
    \centering
    \includegraphics[ width=.5\textwidth,trim={0 0 0 1cm}]{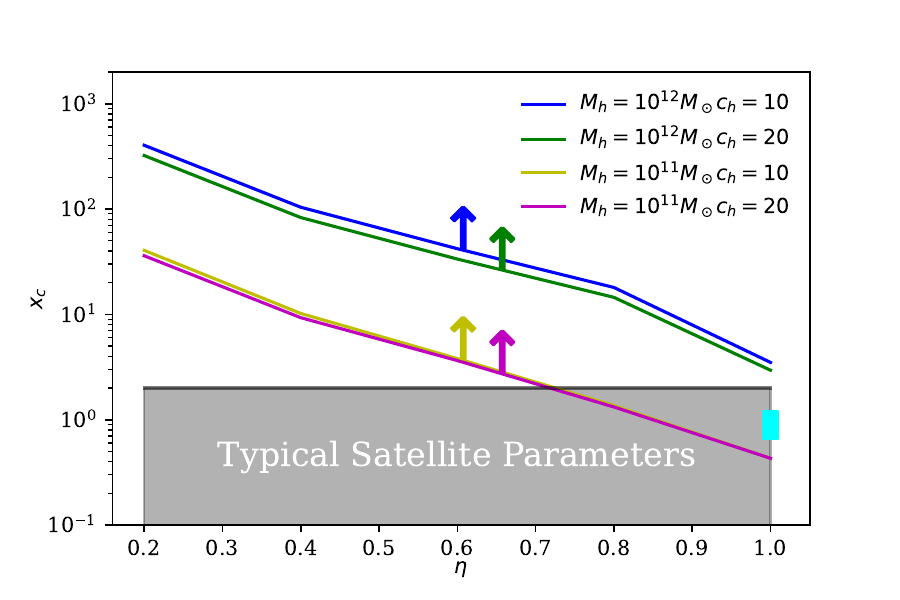}
    \caption{The orbital parameter range, specified by orbital energy $x_c$ and orbit circularity $\eta$, accessible for soliton evolution given our simulations' fiducial box size and resolution and the maximum velocity criterion (Eq.~\ref{vmax}). The black shaded region is typical of CDM subhalos (e.g., see \cite{Ogiya2019}).  The pseudospectral solver we employ, \texttt{UltraDark.jl}, is able to accurately simulate orbits in the regions above the solid lines at our fiducial resolution of 512 grid cells per side and box length of 306 kpc.  These accessible regions depend on the mass and concentration of the host halo, as indicated in the legend. The simulations we run in this work all use $\eta = 1$ (i.e., circular orbits) and $x_c\in [0.7,1.1]$ (see Eq.~\ref{eq:energy}--\ref{eq:circularity}), corresponding to orbital radii between $0.7r_{\text{vir}}$ to $1.1r_{\text{vir}}$ ($67~\text{kpc}$ to $105~\text{kpc}$).  The small cyan region represents the region of parameter space we simulate.}
    \label{DASHparameterspace}
\end{figure}

We initialize our solitons by assuming an equilibrium configuration without self-interactions, with an initial profile following \cite{Edwards2018} (also see \cite{Glennon2020}). Note that, in models with nonzero self-interactions, the solitons do not start in equilibrium; however, for the range of self-interaction models we simulate, oscillations about the equilibrium state are small for isolated versions of our solitons. These small non-equilibrium oscillations do not significantly impact our results. Note that we simulate the evolution of ``bare'' soliton profiles rather than include an outer NFW component of the density profile past some transition radius. We discuss this choice further in Section \ref{sec:results}.

We simulate bare solitons with initial masses ranging from $1.04\times10^8 M_\odot$ to  $1.13\times10^8 M_\odot$, well below the host halo mass of $10^{11} M_\odot$. This allows us to neglect the effects of dynamical friction, which is expected to differ in ULA models relative to CDM (e.g., \cite{Lancaster:2019mde}). Within the narrow range of soliton masses that we simulate, bare solitons survive long enough to complete several orbits, which allows to characterize their disruption timescales accurately, and short enough to disrupt on the order of a Hubble time (see Section \ref{sec:analytic}). We also used a range of values for $x_c$ from 0.7 to 1.1 and a range of $\lambda$ from $-2.4\times10^{-92}$ to $2.4\times10^{-92}$ ($\tilde{\kappa}$ = -5.0 to 5.0). A summary of our simulation parameters is provided in Table~\ref{tab:param_ranges}.

Ideally, we would run simulations with more massive hosts in order to model the effects of tidal stripping of solitons within the Milky Way; however, this is more challenging computationally because the maximum velocity criterion is more easily violated as host mass increases for a given set of orbital parameters. In particular, we use our fiducial box length and grid resolution to calculate the velocity of the orbit based on the host mass and concentration to identify the accessible region of orbital parameter space shown in Fig.~\ref{DASHparameterspace}.  We observe that, as the host mass decreases or the host concentration increases, our simulations are able to access more of the orbital parameter space. This follows because, for any given set of orbital parameters, decreasing the host mass or increasing the host concentration reduces the velocity of the orbit, and therefore makes it less likely that the maximum velocity criterion is violated. Based on Fig.~\ref{DASHparameterspace}, our simulations can robustly probe subhalos of LMC-mass hosts on nearly circular orbits. Subhalos of more massive hosts like the Milky Way, and subhalos on more radial orbits around hosts of mass, remain difficult to simulate robustly given our setup.

\setlength{\tabcolsep}{0pt}
\begin{figure*}[!tbp]
  \centering    
    \begin{tabular}{c}
    \includegraphics[trim=0 0 0 0, width=.7\textwidth]{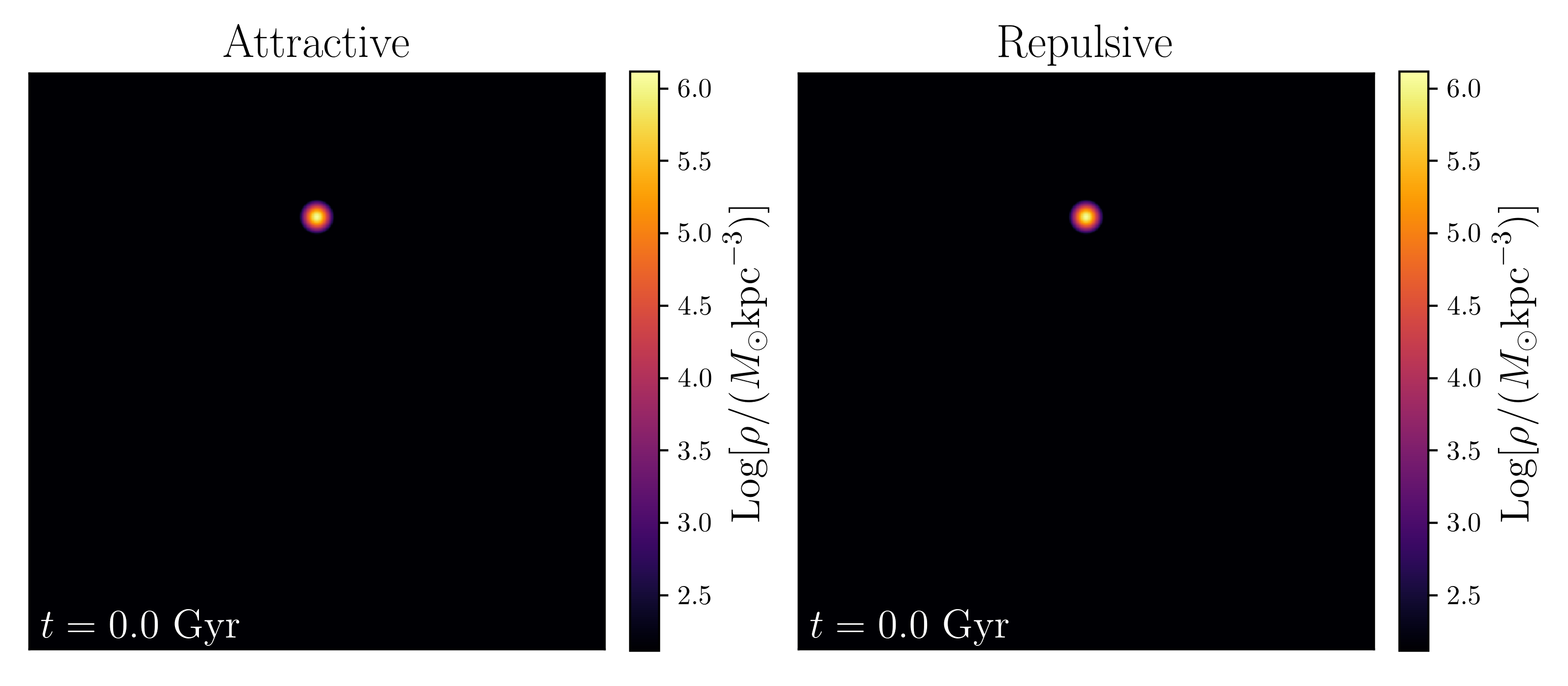}
    \\
    \includegraphics[trim=0 0 0 25, clip, width=.7\textwidth]{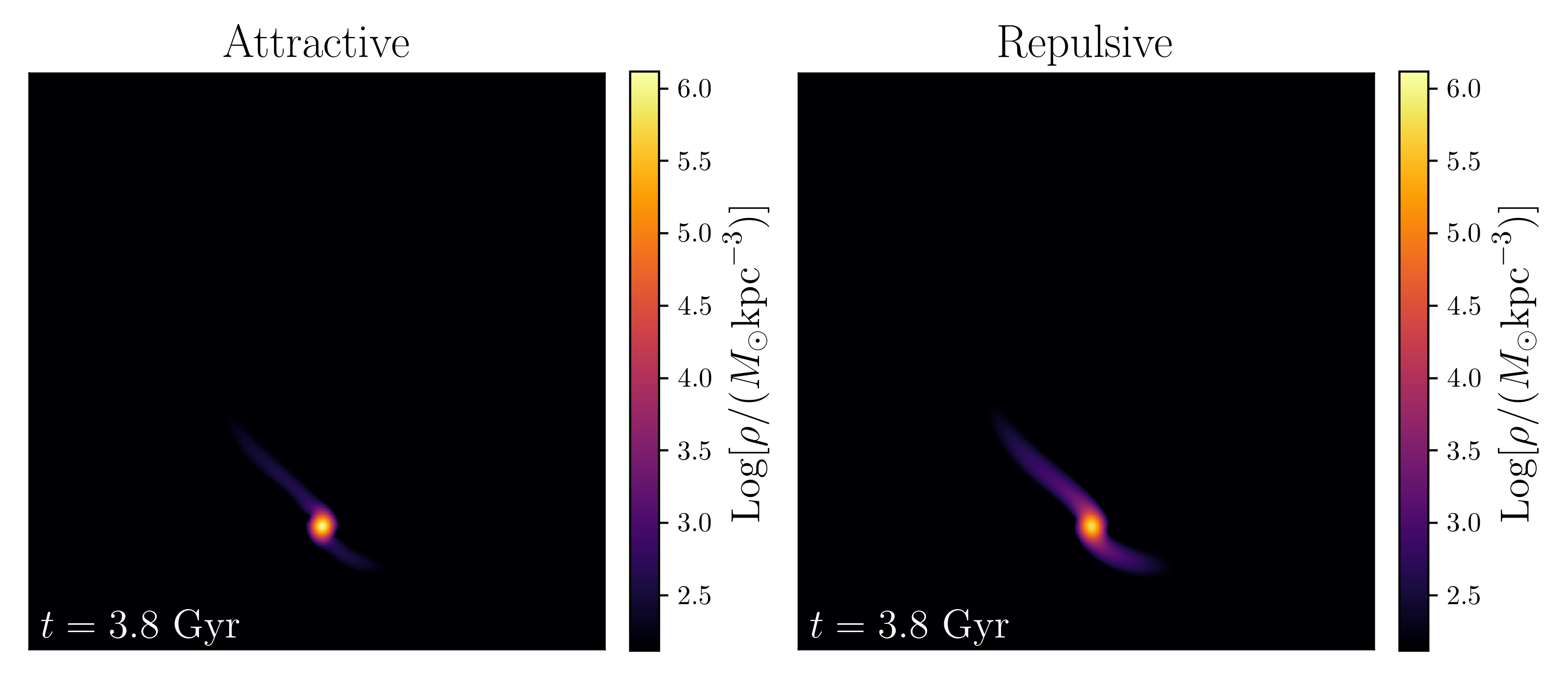}
    \\
    \includegraphics[trim=0 0 0 25, clip, width=.7\textwidth]{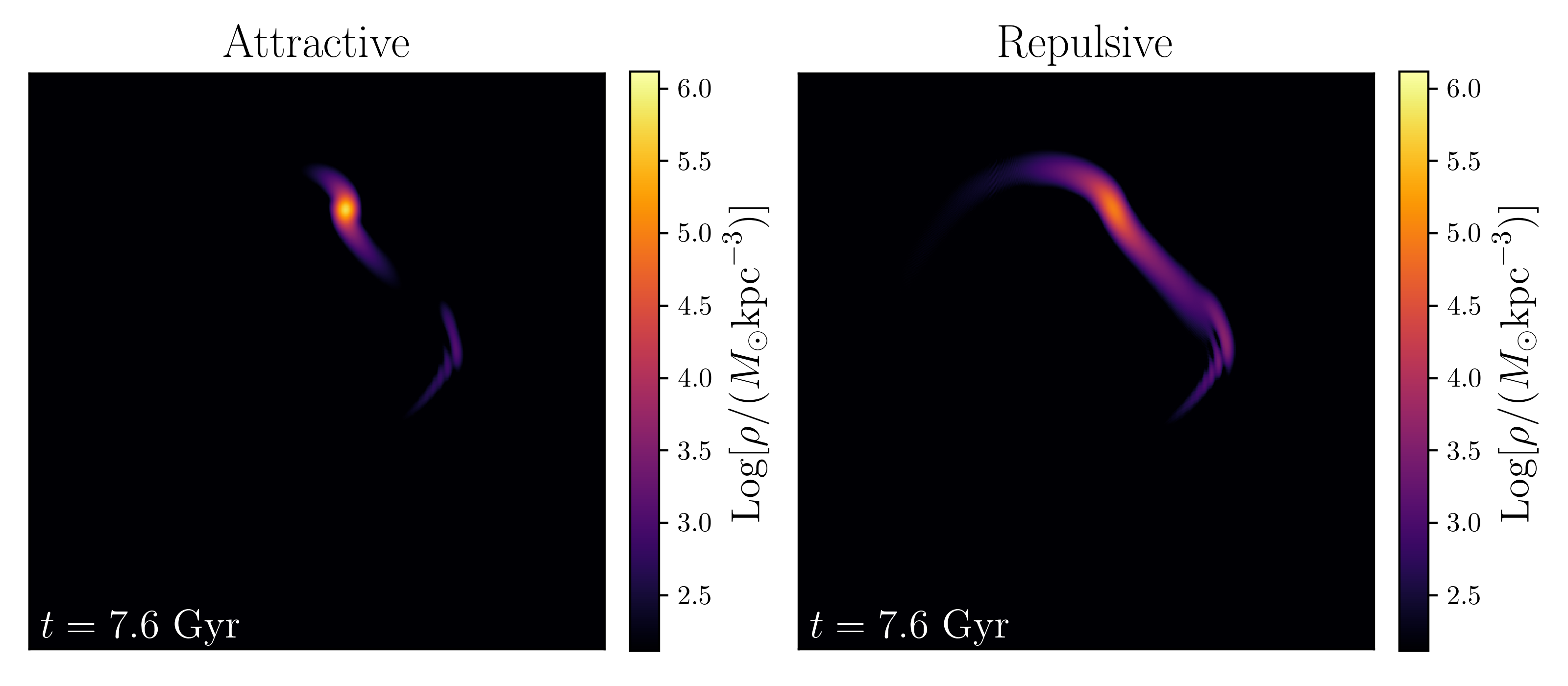}\\
    \includegraphics[trim=0 0 0 25, clip, width=.7\textwidth]{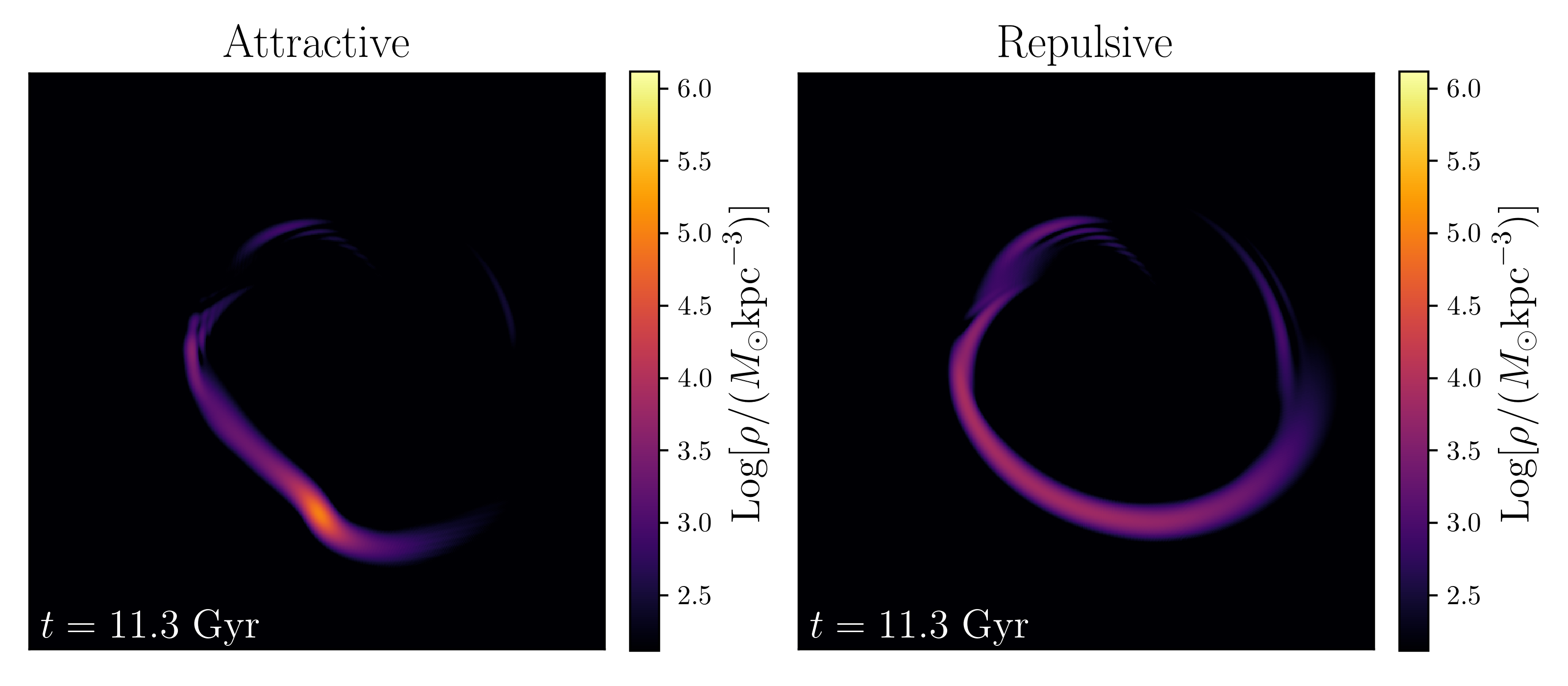}\\
    \end{tabular}
    \caption{%
        Projected density distributions in the orbital plane from representative simulations where a soliton is tidally disrupted in the presence of attractive self-interactions with $\tilde{\kappa} = -5.0$ (left) and repulsive self-interactions with $\tilde{\kappa} = 5.0$ (right). In these simulations, the soliton mass is $1.13 \times 10^8 M_\odot$ and $x_c = 0.8$. The visualized region is approximately 300~kpc long in the x and y directions. Tidal stripping is noticeably more efficient in the case with repulsive self-interactions.  An animation can be found at \href{https://bit.ly/3wzgBQw}{https://bit.ly/3wzgBQw}.
    }
    \label{examlpesim}
\end{figure*}

\begin{figure*}
        \includegraphics[ width=1\textwidth]{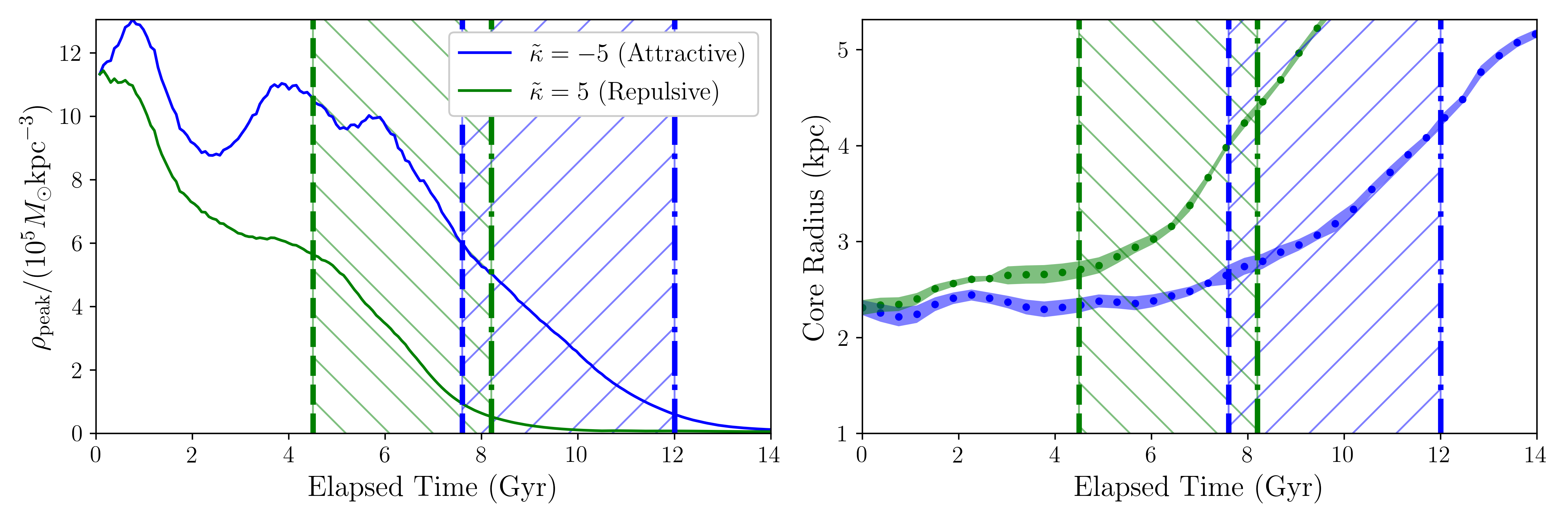}
        \caption{Evolution of the peak density (left panel) and core radius (right panel) for the solitons shown in Fig.~\ref{examlpesim}, for the repulsive (green) and attractive (blue) self-interaction cases. Before tidal disruption sets in, the peak density corresponds to the soliton's core density. The left (right) dashed vertical line represents the $50\%$ ($5\%$) density threshold relative to the initial peak density for each self-interaction model, and the region between the dashed and dot-dashed lines corresponds to the disruption phase. Note that the peak density after the disruption phase is not necessarily located at the center of the remaining soliton core.  In the right panel, the shaded bands represent the maximum uncertainty in the core radius given our spatial resolution and interpolation scheme.  The core radius expands as the soliton is tidally disrupted.
        }
        \label{peakden}
    \end{figure*}

\section{Simulation Results}
\label{sec:results}

We now present the our key results. We begin by briefly describing our initial attempts at simulations, which motivated the choice of soliton masses and orbital configurations presented here.  Next, we describe representative examples of simulations with different self-interaction strengths and signs, and we compare the evolution of the respective solitons as they orbit the central potential. We then summarize how soliton disruption times depend on soliton mass and self-interaction strength and sign. Lastly, we identify a degeneracy between self-interaction strength and soliton mass when calculating the disruption time of solitons.

We reiterate that the simulations we present in this section only include solitonic profiles for the orbiting satellites. In particular, we do not include an outer NFW component of the density profile past some transition radius, which is expected for subhalos in ULA models formed in a cosmological context \cite{Schive2014_2}. This choice allows us to isolate the effects of self-interactions on the disruption time of bare solitons, which disrupt on timescales comparable to the dynamical time set by the central potential (and, thus, on timescales comparable to the Hubble time). We perform and discuss a limited number of simulations where an NFW profile is smoothly added to the soliton in Appendix~\ref{sec:NFW}. These simulations indicate that adding an outer NFW component allows solitons to survive significantly longer, which is expected due to the ``outside-in'' nature of tidal stripping~\cite{vandenBosch2017} and is consistent with previous ULA studies without self-interactions~\cite{Schive:2019rrw}. Nonetheless, we find that the general relation of disruption time dependence on self-interactions that we focus on here is qualitatively unaffected in these scenarios.

In our initial simulations, we used less massive solitons that disrupted nearly instantaneously. An explanation for this can be found in~\cite{Du2018}. Briefly, the ratio of the soliton core density to the average host density within the soliton's orbit is approximately 1.  According to \cite{Du2018}, solitons should disrupt very quickly if the ratio is below 4.5 since this would mean the soliton's core radius is larger than the tidal radius.  We then chose to use higher mass solitons to allow the solitons to survive on cosmological timescales. Thus, our initial tests confirm that bare solitons with low masses are difficult to observe today because they disrupt quickly. Because of this rapid disruption, self-interaction strength and sign had little effect on the results of these simulations. This motivates the range of soliton masses we consider, $1.04$ to $1.13\times 10^{8}~M_{\odot}$, which yield density ratios relative to the host of roughly 50 within the orbital radii we consider, allowing us to study their tidal evolution for significantly longer and to highlight the effects of self-interactions.
    
\setlength{\tabcolsep}{0pt}
\begin{figure*}[!t]
  \centering    
    \includegraphics[width=.9\textwidth]{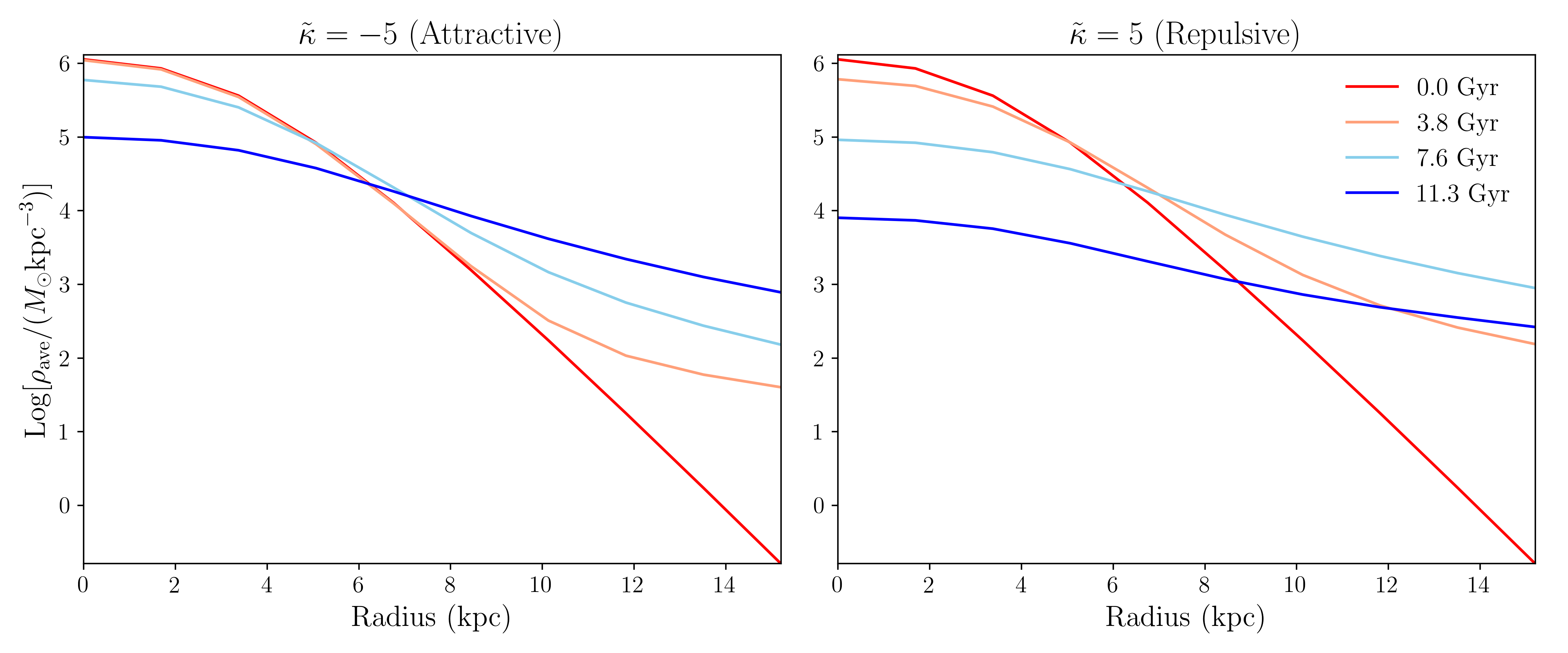}
    \caption{%
        The spherically averaged density profile centered on the soliton, for the simulations in Fig.~\ref{examlpesim}.  The left plot shows the density profile of a soliton where the self-interaction is {\em attractive} and $\tilde{\kappa} = -5.0$. The right plot shows the case where the self-interaction is {\em repulsive} and $\tilde{\kappa} = 5.0$.  We see that the peak density falls faster with time for the simulation where the self-interaction is repulsive.  We also see that, when there are repulsive self-interactions, the overall density profile has more mass in the outer regions at early times. This demonstrates how matter is stripped more quickly in the presence of repulsive self-interactions.
    }
    \label{shellcomparison}
\end{figure*}

\subsection{Examples of Soliton Evolution in the Presence of Self-Interactions}

Two example simulations are shown in Fig.~\ref{examlpesim}.  In these simulations, the soliton mass is set to $1.13\times10^8 M_{\odot}$ and $x_c = 0.8$, corresponding to an orbital radius of $0.8 r_{\text{vir}}\approx 75~\text{kpc}$.  The only difference is that the plots on the left have an attractive coupling of $\tilde{\kappa} = -5.0$ and the plots on the right have a repulsive coupling of $\tilde{\kappa}=5.0$.  In these examples, we can see that the solitonic core is disrupted less quickly (more quickly) in the case with attractive (repulsive) self-interactions. Specifically, Fig.~\ref{peakden} shows the evolution of the peak density and core radius of the solitons from Fig.~\ref{examlpesim}, where the core radius is defined as the distance at which the density drops to $50\%$ of the central density.

The core expands rapidly at late times as the soliton loses mass (recall that soliton size is inversely proportional to soliton mass). In particular, the hatched regions between the vertical lines in Fig.~\ref{peakden} represent the period during which the soliton rapidly disrupts. During this phase, the central density decreases and that the core radius expands less quickly (more quickly) in the case with attractive (repulsive) self-interactions. We study these trends in detail for a range of soliton masses, self-interaction models, and orbital parameters in Section~\ref{sec:disruption_timescale}.

Finally, Fig.~\ref{shellcomparison} shows the spherically averaged density profiles of the two solitons, demonstrating that mass is removed from the central regions more quickly in the case with repulsive self-interactions. This qualitatively explains the differences among the tidal tails in these cases observed in Fig.~\ref{examlpesim}---namely, in the case with attractive self-interactions, the tidal tails are more compact than in the case with repulsive self-interactions; we explore this effect in Appendix \ref{sec:selffriction}.

\subsection{Disruption Timescale}
\label{sec:disruption_timescale}

We quantify the disruption timescale of each simulated soliton by calculating how long it takes for the soliton's peak density to drop to some percentage of its initial value. The peak density is the maximum density found in any grid cell. We used different thresholds for calculating the disruption timescale, corresponding to when the peak density dropped to 50\%, 25\%, 10\%, and 5\% of its initial value. These thresholds are somewhat arbitrary, and are in fact rather extreme in comparison to disruption thresholds typically assumed for CDM subhalos, which lose roughly $99\%$ of their mass once their central density drops by a factor of two \cite{Errani}. Practically, if the threshold is too small, the soliton is no longer well defined when the threshold is reached, such that the peak density over all grid cells no longer corresponds to the center of a coherent object. We have verified that, for the density thresholds we consider, the peak density over all grid cells correctly represents the peak density of the soliton. We measure the disruption timescale using the number of orbits elapsed until the density threshold is reached due to the limited temporal resolution of our simulations.

We find that the relationship between soliton mass and the number of orbits elapsed until disruption is well-described by an exponential relation. A similar relation describes the relationship between the disruption timescale and self-interaction strength. We therefore fit the time elapsed until disruption as a function of initial soliton mass $M_0$ and $\tilde{\kappa}$ according to
\begin{equation}
    S(M,\tilde{\kappa}) = a \exp{\left[(b \tilde{M}_0 + c \tilde{\kappa}_0)/10^2\right]}~\text{Gyr},
    \label{degenfit}
\end{equation}
where $\tilde{M}_0\equiv M_0/10^6\ M_{\mathrm{\odot}}$, and $a$, $b$, and $c$ are dimensionless constants.\footnote{Although this functional form fits our simulation results well, it is not obviously physically motivated. For example, a power-law relation also describes the results of our simulations reasonably well.  Regardless of the fitting function, the degeneracy between self-interaction strength and soliton mass that we present is largely unaffected.}

\begin{figure}[t!]
    \centering
    \includegraphics[trim=10 10 0 10, width=.50\textwidth]{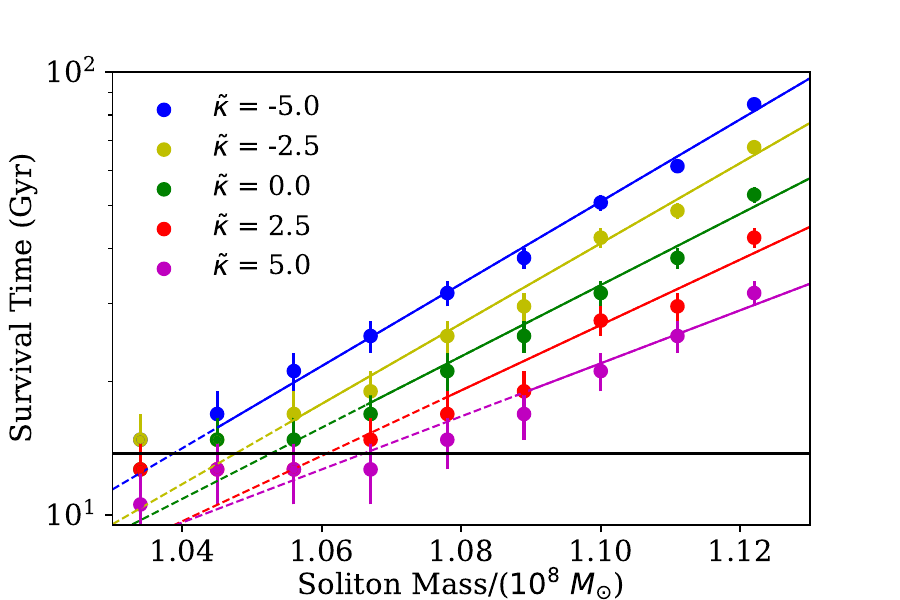}
    \caption{ Dependence of disruption time of a soliton on the soliton mass and self-interaction strength, $\tilde{\kappa}$.  Here, the density cutoff used to determine the disruption time is 50\%, and we assume an orbit where $x_c = 0.9$.  Points of the same color correspond to a specific value of the self-interaction strength, $\tilde{\kappa}$, with solid lines indicating the best fits to these data using  Eq.~\ref{degenfit}. The dashed lines are the extrapolated extensions of the best-fit lines into a region where it is difficult to measure the disruption time accurately. The error bars represent the maximum uncertainty in survival time measurements given our temporal resolution. The black horizontal line indicates the Hubble time. Increasing the soliton mass and making the self-interaction more attractive both cause the soliton to survive longer.}
    \label{survivalvsmass}
\end{figure}

\begin{figure}[b!]
    \centering
    \includegraphics[trim=10 10 0 10, width=.50\textwidth]{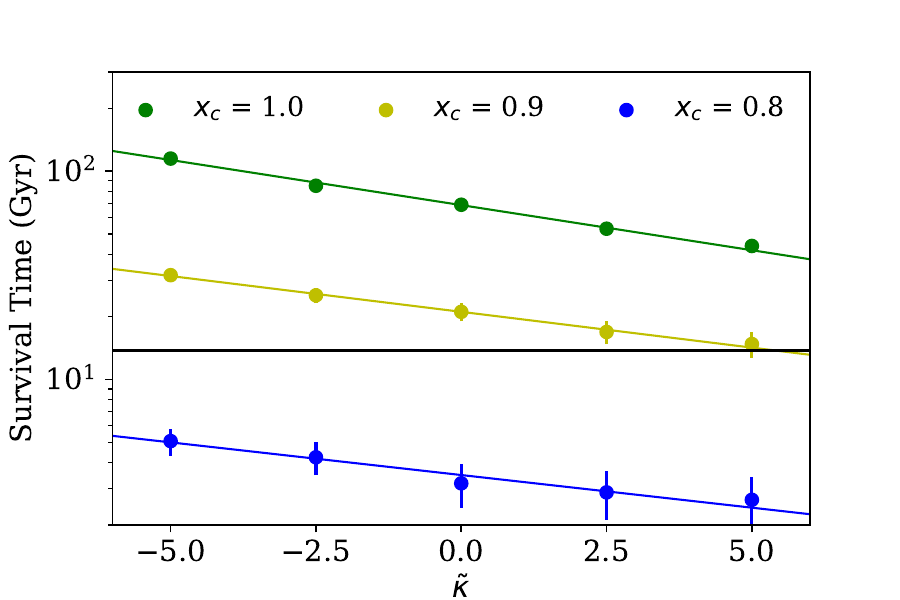}
    \caption{ Dependence of disruption time on self-interaction strength and orbital radius. Here the soliton mass is $1.08 \times 10^8 M_\odot$.  The density cutoff in this plot is 50\%.  Points of the same color represent orbits with the same orbital radius.  The error bars represent the maximum uncertainty in survival time given our temporal resolution.  The black line represents the Hubble time.  Making the self-interaction more attractive and extending the orbital radius both make the soliton more difficult to disrupt.
    }
    \label{survivalvskappa}
\end{figure}

Using Eq.~\ref{degenfit}, we determined the best fit parameters, $a$, $b$, and $c$, that fit our data for different combinations of $x_c$ and density cutoff.  The parameter $a$, can be thought of as the amplitude of the disruption time for our fitting function.  The ratio of parameters $b$ to $c$ describes the degeneracy.  We find that: (1) $a$ becomes larger with decreasing $x_c$ and increasing density thresholds, (2) $b$ becomes larger with increasing $x_c$ and decreases density thresholds, and (3) $c$ becomes more negative with increasing $x_c$ and decreasing density thresholds.

To quantify how the disruption timescale depends on the soliton mass, Fig.~\ref{survivalvsmass} shows the best fit lines using Eq.~\ref{degenfit} for specific values of $\tilde{\kappa}$.  Over the range of masses and $x_c$ we use, a 9\% increase in soliton mass can increase the disruption time by up to a factor of $4.5$ depending on the value of $x_c$ and the density cutoff. This strong dependence on soliton mass results from the sharp transition to runaway disruption below a critical density ratio (see Section \ref{sec:analytic}). We leave a detailed exploration of this dependence across a wider range of soliton masses and self-interaction strengths, and including NFW outskirts (see Appendix \ref{sec:NFW}), to future work.  Similarly, Fig.~\ref{survivalvskappa} shows how the self-interaction strength affects the disruption timescale for a given soliton mass.  Going from $\tilde{\kappa}=5$ to $\tilde{\kappa}=-5$ (getting more attractive), the disruption time increases up to 100\% depending on $x_c$ and the density cutoff.  Both sets of results indicate that Eq.~\ref{degenfit} is a reasonable fit to our simulation results.

\begin{figure}[t!]
    \hspace{-60mm}
    \includegraphics[ width=0.55\textwidth,trim={0 0.75cm 5.25cm 1.75cm}]{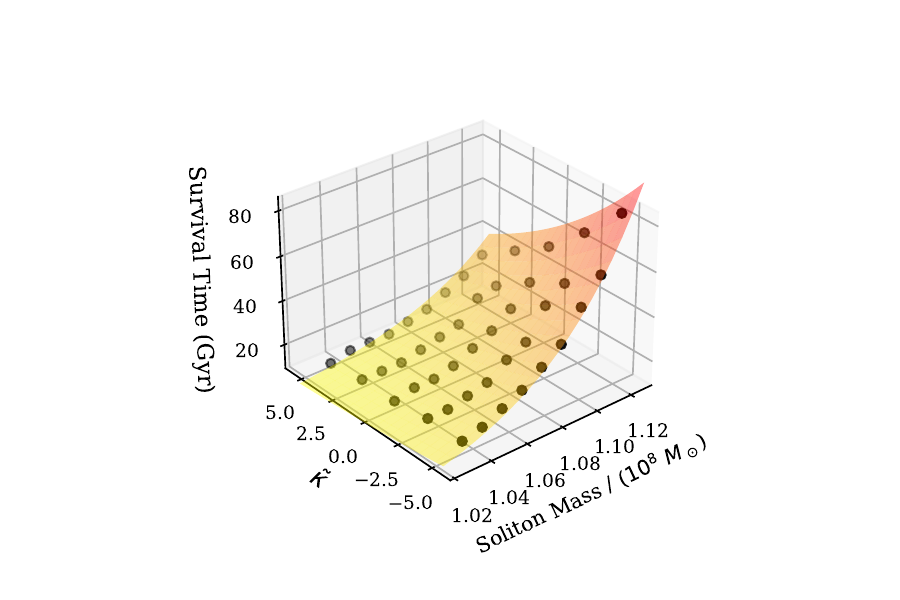}
    \caption{Disruption time as a function of soliton mass and self-interaction strength. Here the orbital energy is $x_c = 0.9$ and the density cutoff is 50\%. The points on the plot represent results from indivudual simulations.  
    The best-fit surface given by Eq.~\ref{degenfit} is shown, where the color represents the disruption time. This demonstrates the degeneracy between different combinations of $\tilde{\kappa}$ and the soliton mass that give the same disruption time. 
    }
    \label{parameterspace}
\end{figure}

These results suggest that there is a degeneracy between soliton mass and self-interaction strength when determining soliton disruption timescales.  In order to gain intuition for this degeneracy, note that attractive (repulsive) self-interactions work to enhance (diminish) the effects of the soliton's own gravity.  In this way, the same disruption timescale can be achieved by making self-interactions more attractive and reducing the soliton mass, or vice versa.  Fig.~\ref{parameterspace} illustrates how different combinations of soliton mass and self-interaction strength can yield the same disruption timescale. The points in Fig.~\ref{parameterspace} are extracted from our simulations, while the contours correspond to the best fit surface generated using Eq.~\ref{degenfit}.

Based on Eq.~\ref{degenfit}, we can estimate the size of the degeneracy between soliton mass and self-interaction strength by requiring that
\begin{equation}
    b \tilde{M}_{0,1} + c\tilde{\kappa}_1 = b \tilde{M}_{0,2} + c\tilde{\kappa}_2,
\end{equation}
which leads to 
\begin{equation}
    r \equiv \frac{b}{c} =  \frac{\tilde{\kappa}_2-\tilde{\kappa}_1}{\tilde{M}_{0,1}-\tilde{M}_{0,2}}.
    \label{rval}
\end{equation}
  
The parameter, $r$, becomes more negative with increasing $x_c$ and increasing density cutoff but has little variation overall.  In particular, $r$ approaches $-2.4$ as the measured disruption time increases (either by extending the density cutoff or increasing $x_c$).  This dependence diminishes as $x_c$ increases.  A summary of these results are found in Table~\ref{fitparams}.

Breaking this degeneracy is important in order to facilitate robust constraints on ULA dark matter, including its potential self-interactions, using observations of small-scale structure at late times. At fixed ULA particle mass, the simplest way to isolate the effects of self-interactions on solitons'  orbital evolution would be to measure the masses of these solitons, or, alternatively, properties of solitons that correlate with soliton mass such as the soliton radius or central density. On the other hand, given a measurement of the self-interaction strength, the soliton mass could be inferred from observational estimates of the disruption timescale for low-mass solitons.

Note that the ULA particle mass can be adjusted independently of soliton mass and self-interaction strength. Thus, we performed several simulations with a particle mass of $m = 2 \times 10^{-22}~\mathrm{eV}$ (recall that our fiducial simulations assume $m=10^{-22}~\mathrm{eV}$). We find that, for a fixed soliton mass, increasing the particle mass increases the disruption timescale coherently across all self-interaction strengths.   This is to be expected as the soliton's central density increases with particle mass, making the soliton more difficult to disrupt.  This does not significantly affect the soliton mass--self-interaction degeneracy reported above.

\section{Discussion}
\label{sec:discussion}

This work serves as a first step toward a systematic study of subhalo and soliton evolution in ULA cosmologies. We now discuss various aspects of this first study that need to be addressed to enable more robust predictions. First, we have mainly considered isolated solitonic profiles herein; as discussed above, a more realistic scenario will be to initialize the central soliton in an extended NFW profile. As indicated by the results in Appendix \ref{sec:NFW}, the presence of additional matter in the outskirts delays the disruption of the central soliton. However, since the self-interactions are most important in the central regions of solitons, the trend with self-interaction strength is likely more robust; we plan to quantify this in detail in future work.

Second, we have not included baryons (or gravitational potentials representing baryonic components) in our simulations, which must be incorporated in a more complete analysis. Baryonic physics could manifest in multiple ways in such a study; for example, the presence of a non-negligible baryonic component can lead to adiabatic contraction in the solitons themselves, potentially making them more resilient to tidal disruption. In analogy to the response of self-interacting dark matter halos to baryons \cite{Fry:2015rta,Elbert160908626,Sameie210212480}, this effect may be enhanced in the presence of ULA self-interactions. The presence of a central baryonic disk in the host halo can also have important effects on disruption times, as has been demonstrated in CDM and SIDM simulations \cite{Garrison-Kimmel170103792,Kelley:2018pdy,Robles:2019mfq}.

\setlength{\tabcolsep}{6pt}
\begin{table}[b!]
\centering
 \begin{tabular}{||c c c||} 
 \hline
 $x_c$ & Density Cutoff & $r$\\ [0.5ex] 
 \hline\hline
 0.7 & 0.50 & -1.5\\
 0.7 & 0.25 & -1.9\\ 
 0.7 & 0.10 & -2.1\\
 0.7 & 0.05 & -2.0\\
 0.8 & 0.50 & -2.0\\
 0.8 & 0.25 & -2.2\\ 
 0.8 & 0.10 & -2.3\\
 0.8 & 0.05 & -2.3\\
 0.9 & 0.50 & -2.3\\
 0.9 & 0.25 & -2.4\\ 
 0.9 & 0.10 & -2.4\\
 0.9 & 0.05 & -2.4\\
 \hline
 \end{tabular}
 \caption{Dependence of the degeneracy parameter (Eq.~\ref{rval}) on the density cutoff and $x_c$.  The ratio $r$ of parameters $b$ to $c$ remains fairly constant.}
 \label{fitparams}
\end{table}

For a full characterization of ULA subhalo and soliton populations in a cosmological context, it is also crucial to sample the evolution and disruption of these systems over the full parameter space of possible orbits, which we were not able to achieve in this study due to resolution limitations (see Fig.~\ref{DASHparameterspace}). Note that this is an issue for all solvers of this type, rather than a limitation specific to \texttt{UltraDark.jl}. There are multiple distinct possible paths forward. One is to continue using static, isolated hosts, while varying the masses and trajectories of the infalling subhalos, as in done in, e.g., \cite{vandenBosch2017,Ogiya2019}. Such an approach is the most straightforward generalization of this study, but to do so, the numerical issues that have been highlighted in Sec.\ \ref{sec:sim_description} must be addressed so that subhalos on orbits with small pericenters can be faithfully modeled. The other approach is to use simulations with cosmological initial conditions, where the evolution of both the host and the infalling subhalos are consistently tracked over the entire history. Within such an approach, it is possible to either zoom in on the evolution of particular hosts with very high resolution, or consider the (lower-resolution) subhalo populations of all hosts in a chosen mass range within the simulation volume. We will address these issues in future work.

One of the central results of the present work is characterizing the degeneracy between the soliton mass and the self-interaction strength of the ULA in terms of disruption time in a central potential. This has deep implications for the use of the subhalo populations as a probe of ultralight dark matter. In particular, analyzing data without accounting for the possible presence of self-interactions can bias inferences about the mass of the central soliton, and in turn the particle mass of an ULA dark matter candidate. Further, this suggests that to disentangle the effects of the soliton mass and the self-interaction strength, we need to consider---in addition to subhalo abundances and density profiles---other observables that do not share the same degeneracy direction. For example, self-interactions may affect structure formation at early times, as encoded in the linear matter power spectrum (e.g., \cite{Shapiro:2021hjp}), relative to a corresponding ULA model with the same particle mass but without self-interactions. Complementary cosmological observables including the small-scale matter power spectrum, cluster mergers \cite{Fan2016}, the expansion history and number of relativistic degrees of freedom in the early universe \cite{Li:2013nal}, and the primordial gravitational wave background \cite{Li:2016mmc} may further differentiate ULA models with and without self-interactions.

\section{Conclusion}
\label{sec:conclusions}

In this work, we have shown how self-interactions in an ultralight dark matter model are capable of affecting soliton disruption times.  We have done this by running simulations in \texttt{UltraDark.jl} that models a $\approx 10^8 M_{\odot}$ soliton in a circular orbit around a $10^{11} M_{\odot}$ host.  Our results indicate that  self-interactions can significantly impact soliton disruption timescales at fixed ultra-light axion (ULA) particle mass and soliton mass for astrophysically relevant orbital configurations.  Our main results are summarized below.
\begin{itemize}
    \item We find that ULA self-interaction strength and sign affect soliton disruption, such that solitons disrupt more (less) efficiently for repulsive (attractive) self-interactions. 
    \item For plausible variations in the dimensionless ULA self-coupling $\tilde{\kappa}$, soliton disruption timescales change by about 30\% at fixed soliton mass and orbtal configuration.
    \item We identify a degeneracy between the self-interaction strength and the soliton mass in determining the disruption time.
    \item Bare solitons with a central density below a certain threshold of the host halo's density within the orbital radius disrupt in $\ll 1~\text{Gyr}$, even with relatively strong attractive self-interactions, corroborating previous results (e.g., \cite{Du2018}).
\end{itemize}

Our results indicate that the effects of ULA self-interactions should be accounted for to accurately model solitons' nonlinear evolution within larger host halos. Conversely, neglecting the effects of self-interactions may lead to biased inferences on the properties of solitons and on the ULA mass. Joint inferences of the ULA particle mass and self-interactions therefore represent an important area for future work. Furthermore, our results qualitatively suggest that the internal structure of surviving solitons can be affected by self-interactions, which may alter the density profiles and core--halo mass relations assumed when fitting inferred dwarf galaxy density profiles (e.g., \cite{Bar:2021kti,Hayashi:2021xxu})

These considerations are particularly important given imminent advances in the precision of small-scale structure measurements. For example, over the next two decades, the Vera C.\ Rubin Observatory and the Nancy Grace Roman Space Telescope are expected to dramatically expand the populations of stellar streams, ultra-faint dwarf galaxies, and strong gravitational lenses available for detailed follow-up studies. In this way, these facilities will precisely measure the abundance and properties of small halos and subhalos on scales that originally motivated ULA dark matter \cite{LSST:2008ijt,Roman,Drlica-Wangner2019,Bechtol2019,Gezari:2022rml}.  Our work takes a new step towards constructing a robust model of small-scale structure in realistic ULA particle models, which will be necessary to interpret these exciting observations.

\acknowledgments

We thank Andrew Eberhardt, Phil Mansfield, Katelin Schutz, and Richard Easther for comments on the manuscript, and we thank JiJi Fan and Luna Zagorac for helpful discussions.

This research made use of computational resources at SLAC National Accelerator Laboratory, a U.S.\ Department of Energy Office; the authors are thankful for the support of the SLAC computing team and all administrative and custodial staff at our respective institutions. This research was supported in part by the National Science Foundation under Grants No. NSF PHY-1748958 and No. 1929080. CPW thanks the late Karsten Pohl for actively supporting the application for NSF grant No. 1929080.  This work was performed in part at Aspen Center for Physics, which is supported by National Science Foundation under Grant No. PHY-1607611.  This work was partially supported by a grant from the Simons Foundation.  This work was partially supported by a grant from the Sloan Foundation.

\appendix

\section{Resolution Tests}

\begin{figure}[b!] 
    \hspace{-8mm}
    \includegraphics[ width=0.5\textwidth]{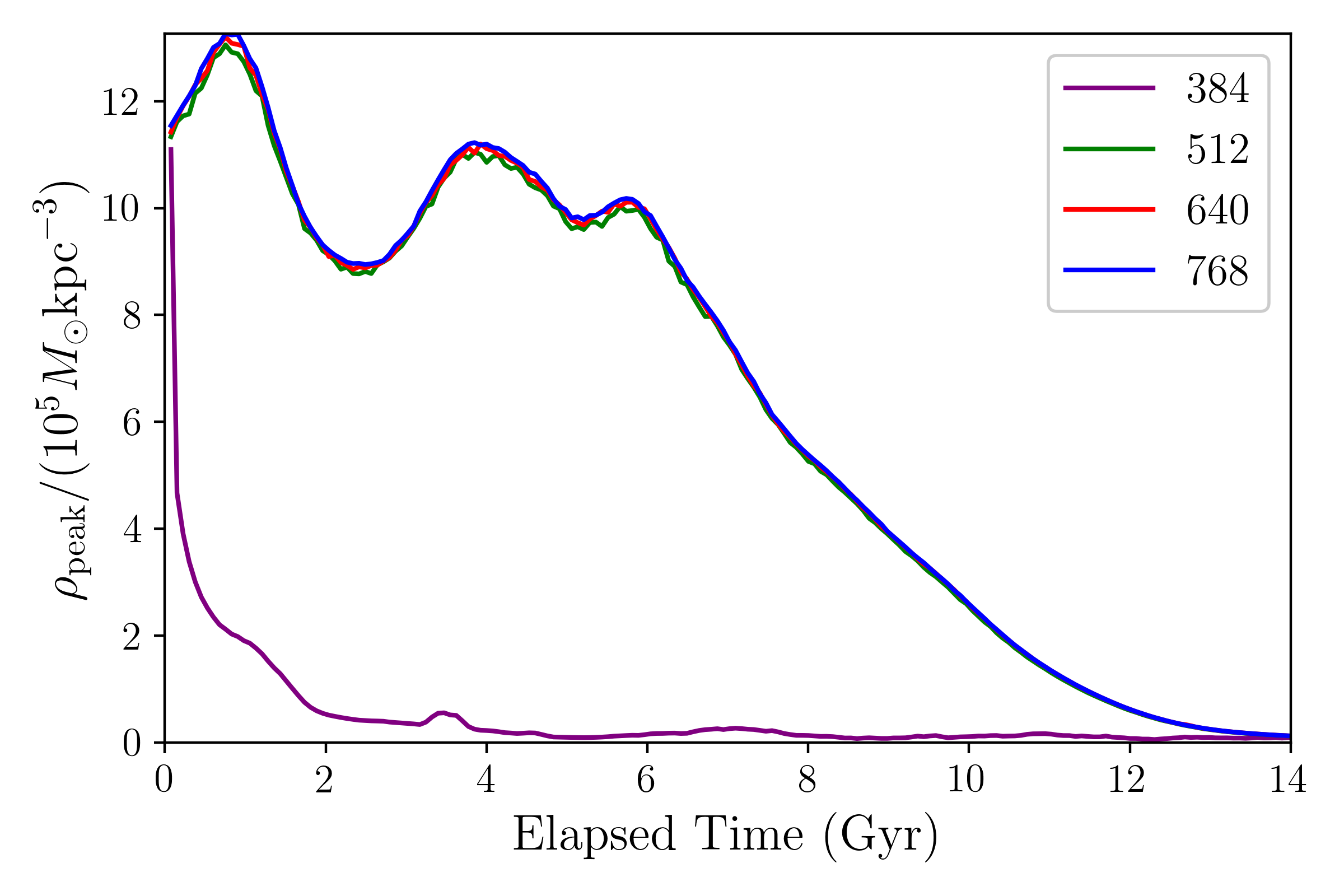}
    \caption{The evolution of soliton peak density for different simulation resolutions, labeled by the number of grid cells per side.  All simulation shown here use $M_{\text{sol}} = 1.13\times 10^{8}~M_{\odot}$, $x_c = 0.8$, and $\tilde{\kappa} = -5.0$.  The disruption times calculated at our fiducial resolution of 512 grid cells per side are stable at the $1\%$ level for higher-resolution runs. Note that the peak density cannot be calculated meaningfully at a resolution of 384 grid cells per side because the maximum velocity criterion (Eq.~\ref{vmax}) is violated for this soliton and orbital configuration.}
    \label{restest}
\end{figure} 

In order to verify that the disruption times we calculated do not change significantly at resolutions higher than our fiducial choice of 512 grid cells per side, we performed several simulations at varying resolutions.  For these tests, we adopt a soliton mass of $1.13\times 10^{8}~M_{\odot}$, $x_c = 0.8$, and two self-interaction models: $\tilde{\kappa} = -2.5$ and $\tilde{\kappa} = -5$.  We chose these parameters because the disruption times at our fiducial resolution were relatively short. In particular, since the simulation run time scales as the number of grid cells per side cubed, simulations with very large disruption times take an infeasible amount of computing resources for very high-resolution runs.

\begin{figure}[b!]
        \hspace{-8mm}
        \includegraphics[ width=.5\textwidth]{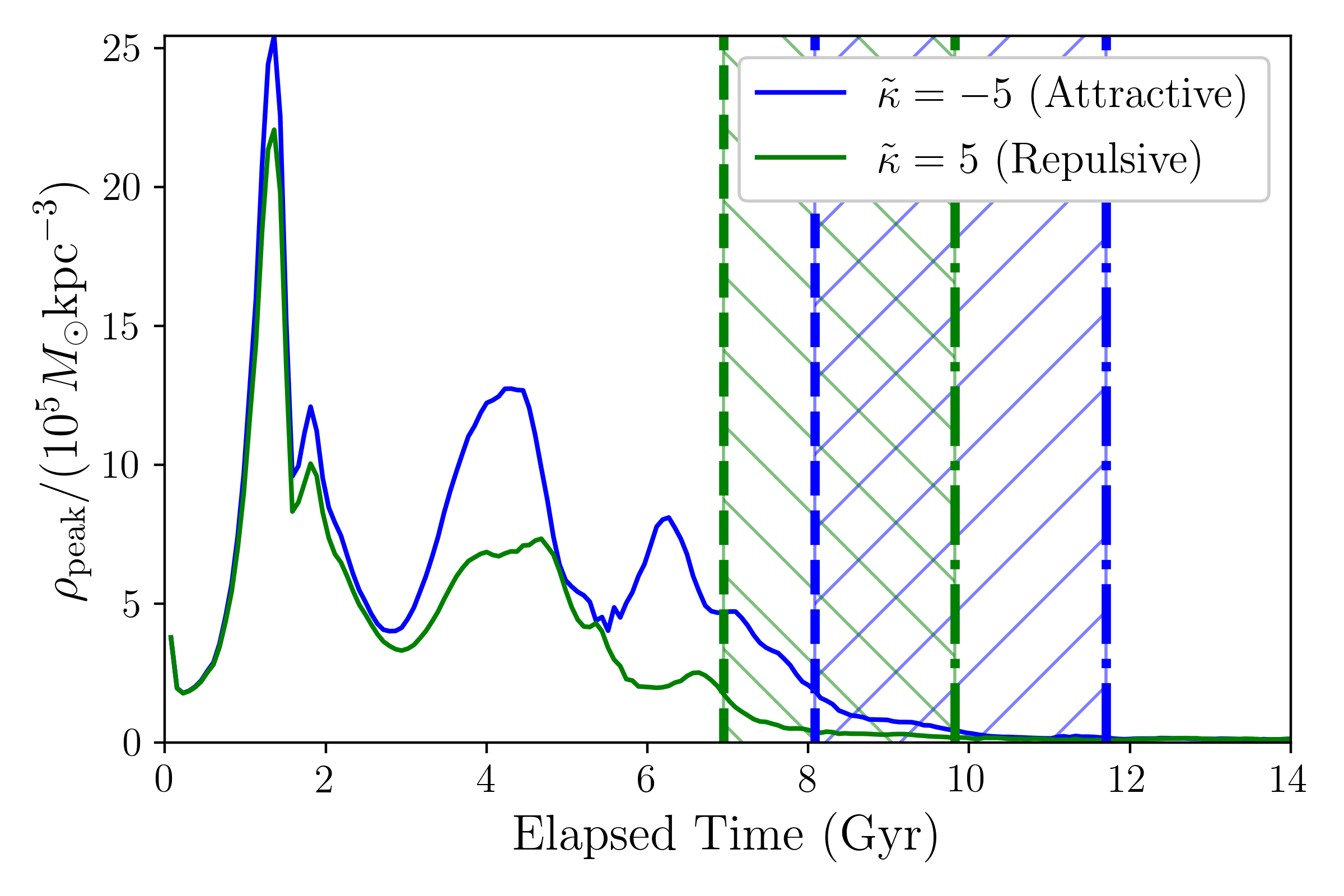}
        \caption{The evolution of the peak density for a combined soliton-NFW profile in the presence of attractive (blue) versus repulsive (green) self-interactinos. These simulations use $x_c = 0.8$ and a soltion mass of $5.5\times 10^7 M_{\odot}$. The dashed (dot-dashed) vertical line represents the $50\%$ ($5\%$) density threshold relative to the initial peak density. For each self-interaction model, the region between the dashed and dot-dashed lines corresponds to the disruption phase.
        }
      \label{peakdennfw}
    \end{figure}

We tested these simulations at resolutions of 384, 512, 640, an 768 grid cells per side.  At a resolution of 384, the maximum velocity criterion (Eq.~\ref{vmax}) was violated and no meaningful disruption timescale could be calculated. At resolutions of 512 and above, we found that the calculated disruption time differed by at most $1\%$.  Fig.~\ref{restest} demonstrates this by showing the evolution of soliton peak density for the resolutions we tested. Thus, we conclude that our fiducial results are not significantly impact by resolution.

\section{Adding an NFW Region}
\label{sec:NFW}

In cosmological simulations of ULAs, subhalos are expected to form with a solitonic core and an NFW outer region \cite{Schive2014,Marsh2015a}. Previous studies indicate that adding an outer NFW region around a central soliton causes a tidally evolving soliton to survive for significantly longer than ``bare'' solitons \cite{Schive:2019rrw}. We test this in our simulations by adding an outer profile to our solitons using the density profile
\begin{equation}
    \rho_{\mathrm{NFW}}(r) = \frac{\rho_0}{\frac{r}{r_s}\left(1+\frac{r}{r_s}\right)^2}.
\end{equation}
Here, $r_s$ is the scale radius of the NFW profile and $\rho_0$ is given by
\begin{equation}
    \rho_0 = \rho_{\text{sol}}(r_t)\frac{r_t}{r_s}\left(1+\frac{r_t}{r_s}\right)^2,
\end{equation}
where $r_t$ is the transition radius and $\rho_{\text{sol}}$ is the soliton's density at the transition radius. 
    
\begin{figure*}[t!]
    \hspace{-8mm}
    \includegraphics[ width=.95\textwidth]{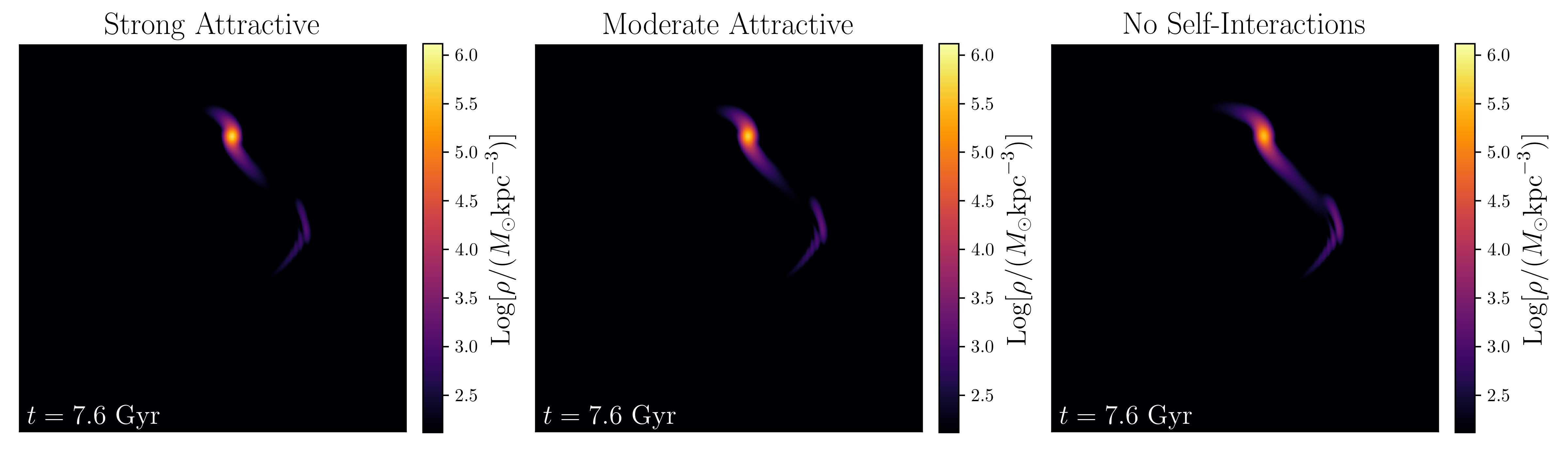}
    \caption{Projected density distributions in the orbital plane from simulations where a soliton is tidally disrupted in the presence of attractive self-interactions with $\tilde{\kappa} = -5.0$ (left) and $\tilde{\kappa} = -2.5$ (center), and with no self-interactions (right)  Note that the second density peak (i.e., the tidal tail of the stripped soliton) is increasingly coherent for stronger attractive self-interactions.  In each simulation, the soliton mass is $1.13\times10^8 M_{\odot}$ and $x_c = 0.8$.}
    \label{self-friction}
\end{figure*} 

We added this profile to the solitonic profile two different ways.  First, we added the profiles together in a piecewise fashion, with a transition from soliton to NFW profiles at $r_t$.  Adding the profiles in this way created a discontinuity in the slope of the density profile at the transition radius.  As a result of this discontinuity, the NFW region in these simulations appeared to separate from the soliton. To alleviate this effect, we performed another set of simulations where the soliton and NFW profiles were simply added together. In this case, the inner regions are dominated by the soliton profile and the outer regions are dominated by the NFW profile.  This method created a smooth profile while increasing the total mass in the inner regions.

We ran several simulations with this combined profile with different self-interactions and found that adding the NFW region increased the lifetimes of the solitons in each scenario. We also found that the qualitative effects of self-interactions on soliton's disruption times remained unchanged.  In particular, at fixed soliton mass and orbital parameters, we find the solitons survive longer in the presence of attractive self-interactions in comparison to models with with repulsive self-interactions even when an outer NFW region is included in our simulations.

As an example, the evolution of the peak density for a set of sample simulation is in Fig.~\ref{peakdennfw}.  Note that in these simulations, we use a smaller soliton mass to reduce the disruption time because adding the NFW region extends the soliton lifetime significantly. In this example simulation, we find that disruption timescales are increased by $\sim 50\%$ for ULA models with attractive versus repulsive self-interactions. In comparison, our fiducial simulations with larger soliton mass and no NFW region exhibit a factor of $\sim 2$ difference among disruption timescales in attractive versus repulsive self-interaction models (see Fig.~\ref{peakden}). Exploring how the impact of self-interactions depends on the entire extent of the soliton profile over a wider range of soliton masses is therefore an interesting avenue for future work.

In Fig.~\ref{peakdennfw}, we also see that the shapes of the peak density curves look different than those in Fig.~\ref{peakden}.  This is caused by how the initial profiles were initialized.  Without the NFW region, the soliton starts out close to equilibrium.  When we include the NFW region, the system starts much further from equilibrium.  The added mass in the central region causes the soliton/NFW profile to collapse initially, and explains the sharp increase in the maximum density.  After this, the soliton/NFW profile oscillates in size until it disrupts.

\section{Tidal Tails of Stripped Solitons}
\label{sec:selffriction}

In our simulations, gaps often appear in the tails of solitons' stripped matter (e.g., see Fig.~\ref{examlpesim}). These tidal tails result in ``self-friction,'' i.e., they apply a torque to the main body of the soliton, causing its rotation to slow~\cite{Miller2020}. Interestingly, the clumpiness of the tidal tails is often enhanced for simulations performed with attractive self-interactions. This effect can be seen in Fig.~\ref{self-friction}, which demonstrates that, as self-interactions become more attractive, the second density peak associated with the soliton's tidal tail becomes more coherent.

\bibliography{masterlibrary}

\end{document}